\documentclass[10pt]{sig-alternate-05-2015}
\pdfoutput=1 
\usepackage[T1]{fontenc}
\usepackage{subfigure}
\usepackage{multirow}
\usepackage{multicol}

\begin{document}


\CopyrightYear{2016} 
\setcopyright{acmlicensed}
\conferenceinfo{IMC 2016,}{November 14 - 16, 2016, Santa Monica, CA, USA}
\isbn{978-1-4503-4526-2/16/11}\acmPrice{\$15.00}
\doi{http://dx.doi.org/10.1145/2987443.2987472}

\clubpenalty=10000 
\widowpenalty = 10000





%


\title{A First Look at Quality of Mobile Live Streaming Experience: the Case of Periscope}


%
%
%
%
%

\numberofauthors{3} 
%
\author{
%
%
\alignauthor
Matti Siekkinen\\
       \affaddr{School of Science}\\
       \affaddr{Aalto University, Finland}\\
       \email{matti.siekkinen@aalto.fi}
\alignauthor
Enrico Masala\\
       \affaddr{Control \& Comp. Eng. Dep.}\\
       \affaddr{Politecnico di Torino, Italy}\\
       \email{masala@polito.it}
\alignauthor
Teemu K\"am\"ar\"ainen\\
       \affaddr{School of Science}\\
       \affaddr{Aalto University, Finland}\\
       \email{teemu.kamarainen@aalto.fi}
}
\date{May 3, 2016}

\maketitle
\begin{abstract}
  Live multimedia streaming from mobile devices is rapidly gaining
  popularity but little is known about the QoE they provide.
  In this paper, we examine the Periscope service. We first crawl the
  service in order to understand its usage patterns. Then, we study
  the protocols used, the typical quality of experience indicators,
  such as playback smoothness and latency, video quality, and the
  energy consumption of the Android application.
\end{abstract}

%
%


%
%

%
%


\keywords{Mobile live streaming; QoE; RTMP; HLS; Periscope}

\section{Introduction}
\label{sec:intro}

Periscope and Meerkat are services that enable users to broadcast live
video to a large number of viewers using their mobile device. They
both emerged in 2015 and have since gained popularity fast. Periscope,
which was acquired by Twitter before the service was even launched,
announced in March 2016 on their one year birthday that over 110 years
of live video was watched every day with the
application~\cite{pcope_oneyear}. Also Facebook has recently launched
a rival service called Facebook Live.

Very little details have been released about how these streaming
systems work and what kind of quality of experience (QoE) they
deliver. One particular challenge they face is to provide low latency
stream to clients because of the features that allow feedback from
viewers to the broadcaster in form of a chat, for example. Such
interaction does not exist with ``traditional'' live video streaming
systems and it has implications on the system design (e.g., protocols
to use).

We have measures the Periscope service in two ways. We first created a
crawler that queries the Periscope API for ongoing live streams and
used the gathered data of about 220K distinct broadcasts to analyze
the usage patterns. Second, we automated the process of viewing
Periscope broadcasts with an Android smartphone and generated a few
thousand viewing sessions while logging various kinds of data. Using
this data we examined the resulting QoE. In addition, we analyzed the
video quality by post processing the video data extracted from the
traffic captures. Finally, we studied the application induced energy
consumption on a smartphone.

Our key findings are the following: 1) 2 Mbps appears to be the key
boundary for access network bandwidth below which startup latency and
video stalling clearly increase. 2) Periscope appears to use the HLS
protocol when a live broadcast attracts many participants and RTMP
otherwise. 3) HLS users experience a longer playback latency for the
live streams but typically fewer stall events. 4) The video bitrate
and quality are very similar for both protocols and may exhibit
significant short-term variations that can be attributed to extreme
time variability of the captured content. 5) Like most video apps,
Periscope is power hungry but, surprisingly, the power consumption
grows dramatically when the chat feature is turned on while watching a
broadcast. The causes are significantly increased traffic and elevated
CPU and GPU load.

\section{Methods and Data Collection}
\label{sec:}

The Periscope app communicates with the servers using an API that
is private in that the access is protected by SSL. To get access to
it, we set up a so called SSL-capable \textit{man-in-the-middle
  proxy}, i.e. mitmproxy~\cite{mitmproxy}, in between the mobile
device and the Periscope service as a transparent proxy.
The proxy intercepts the HTTPS requests sent by the mobile device and
pretends to be the server to the client and to be the client to the
server.
The proxy enables us to examine and log the exchange of requests and
responses between the Periscope client and servers. The Periscope iOS
app uses the so called \textit{certificate pinning} in which the
certificate known to be used by the server is hard-coded into the
client. Therefore, we only use the Android app in this
study.

We used both Android emulators (Genymotion~\cite{genymotion}) and smartphones in
the study. We generated two data sets.
For the first one, we used an Android emulator and developed an
\textit{inline script} for the mitmproxy that crawls through the service
by continuously querying about the ongoing live broadcasts.
The obtained data was used to analyze the usage patterns
(Sec.~\ref{sec:usage}).

The second dataset was generated for QoE analysis (Sec.~\ref{sec:qoe})
by automating the broadcast viewing process on a smartphone. The app
has a ``Teleport'' button which takes the user directly to a randomly
selected live broadcast.
Automation was achieved with a script that sends tap events through
Android debug bridge (adb) to push the Teleport button, wait for 60s,
push the close button, push the ``home'' button and repeat all over
again. The script also captures all the video and audio traffic using
tcpdump. Meanwhile, we ran another inline script with mitmproxy that
dumped for each broadcast viewed a description and playback
statistics, such as delay and stall events, which the application
reports to a server at the end of a viewing session. It is mainly
useful for those streaming sessions that use the RTMP protocol because
after an HTTP Live Streaming (HLS) session, the app reports only the
number of stall events. We also reconstruct the video data of each
session and analyze it using a variety of scripts and tools.  After
finding and reconstructing the multimedia TCP stream using
wireshark~\cite{wireshark}, single segments are isolated by saving the
response of HTTP GET request which contains an MPEG-TS
file~\cite{mpeg_ts_standard} ready to be played.  For RTMP, we exploit
the wireshark dissector which can extract the audio and video
segments.   
The libav~\cite{libav} tools have been used to
inspect the multimedia content and decode the video in full for the
analysis of Sec.~\ref{sec:audio_video_quality}.

In the automated viewing experiments, we used two different phones:
Samsung Galaxy S3 and S4. The phones were located in Finland and
connected to the Internet by means of reverse tethering through a USB
connection to a Linux desktop machine providing them with over 100Mbps
of available
bandwidth both up and down stream. In
some experiments, we imposed artificial bandwidth limits with the
\texttt{tc} command on the Linux host. For latency measurement
purposes (Section~\ref{sec:stall_latency}), NTP was enabled on the
desktop machine and used the same server pool as the Periscope app.

\section{Periscope Overview}
\label{sec:periscope}

Periscope enables users to broadcast live video for other users to
view it. Both public and private broadcasting is available. Private
streams are only viewable by chosen users. 
Viewers can use text chat and emoticons to give feedback to the
broadcaster. The chat becomes full when certain number of viewers have
joined after which new joining users cannot send messages. Broadcasts
can also be made available for replay.

A user can discover public broadcasts in three ways: 1) The app
shows a list of about 80 ranked broadcasts in addition to a couple of
featured ones. 2) The user can explore the map of the world in
order to find a broadcast in a specific geographical region. 
The map shows only a fraction of the broadcasts available in a
large region and more broadcasts become visible as the user zooms
in. 3) The user can click on the ``Teleport'' button to start
watching a randomly selected broadcast.

\begin{table}[t]
\caption{Relevant Periscope API commands.}
\begin{center}
\label{tab:pcope_api}
 {\scriptsize
\begin{tabular}{|l||p{2.5cm}|p{2.5cm}|}
\hline
\textbf{API request}  & \textbf{request contents} & \textbf{response contents}\\
\hline \hline
{\tt mapGeoBroadcastFeed}  & Coordinates of a rectangle shaped geographical area & List of broadcasts located
inside the area\\ \hline
{\tt getBroadcasts}  & List of 13-character broadcast IDs & Descriptions of
broadcast IDs (incl. nb of viewers)\\ \hline
{\tt playbackMeta}  & Playback statistics & nothing \\ \hline
\end{tabular}
}
\end{center}
\end{table}

Since the API is not public, we examined the HTTP requests and
responses while using the app through the mitmproxy in order to
understand how the API works. The application communicates with the
servers by sending POST requests containing JSON encoded attributes to
the following address:
\url{https://api.periscope.tv/api/v2/apiRequest}. The
\texttt{apiRequest} and its contents vary according to what the
application wants to do. Requests relevant to this study are listed in
Table~\ref{tab:pcope_api}.

Periscope uses two kinds of protocols for the video stream delivery:
Real Time Messaging Protocol (RTMP) using port 80 and HTTP Live
Streaming (HLS) because RTMP enables low latency
(Section~\ref{sec:qoe}), while HLS is employed to meet scalability
demands. Also Facebook Live uses the same set of
protocols\cite{fblive}.
Further investigation reveals that the RTMP streams are always delivered
by servers running on Amazon EC2 instances. For example, the IP
address that the application got when resolving
\url{vidman-eu-central-1.periscope.tv} gets mapped to
\url{ec2-54-67-9-120.us-west-1.compute.amazonaws.com} when performing
a DNS reverse lookup. In contrast, HLS video segments are delivered by
Fastly CDN. RTMP streams use only one connection, whereas HLS may
sometimes use multiple connections to different servers in parallel to fetch the
segments, possibly for load balancing and/or resilience reasons. We study
the logic of selecting the protocol and its impact on user experience
in Section~\ref{sec:qoe}. Public streams are delivered using plaintext
RTMP and HTTP, whereas the private broadcast streams are encrypted
using RTMPS and HTTPS for HLS. The chat uses Websockets to deliver
messages.

\section{Analysis of Usage Patterns}
\label{sec:usage}

We first wanted to learn about the usage patterns of Periscope.
The application does not provide a complete list of broadcasts and the
user needs to explore the service in ways described in the previous
section. In late March of 2016, over 110 years
of live video were watched every day through
Periscope~\cite{pcope_oneyear}, which roughly translates into 40K live
broadcasts ongoing all the time.

We developed a crawler by writing a mitmproxy inline script that
exploits the \texttt{/mapGeoBroadcastFeed} request of the Periscope
API. The script intercepts the request made by the application after
being launched and replays it repeatedly in a loop with modified
coordinates and writes the response contents to a file. It also sets
the \texttt{include\_replay} attribute value to false in order to only
discover live broadcasts. In addition, the script intercepts
\texttt{/getBroadcasts} requests and replaces the contents with the
broadcast IDs found by the crawler since previous request and extracts
the viewer information from the response to a file. 

We faced two challenges: First, we noticed that when specifying a
smaller area, i.e. when user zooms in the map, new broadcasts are
discovered for the same area. Therefore, to find a large fraction of
the broadcasts, the crawler must explore the world using small enough
areas.
Second, Periscope servers use rate limiting so that too frequent
requests will be answered with HTTP 429 (``Too many requests''), which
forces us to pace the requests in the script and increases the
completion time of a crawl. If the crawl takes a long time, it will
miss broadcast information.

\begin{figure}[t!]
\begin{center}
\subfigure[absolute]{\includegraphics[width=0.49\linewidth]{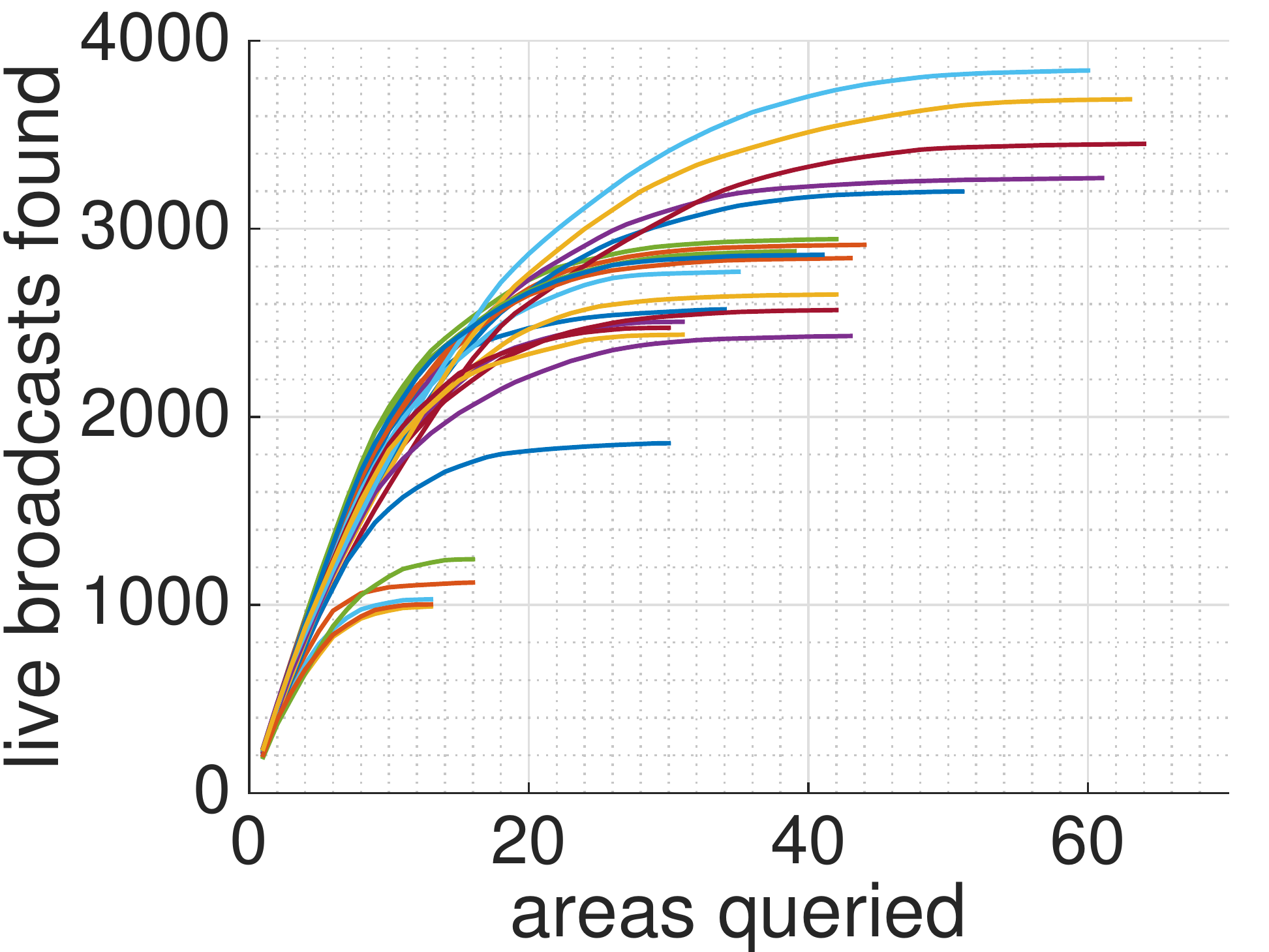}}
\subfigure[relative]{\label{fig:deep_crawl_rel}\includegraphics[width=0.49\linewidth]{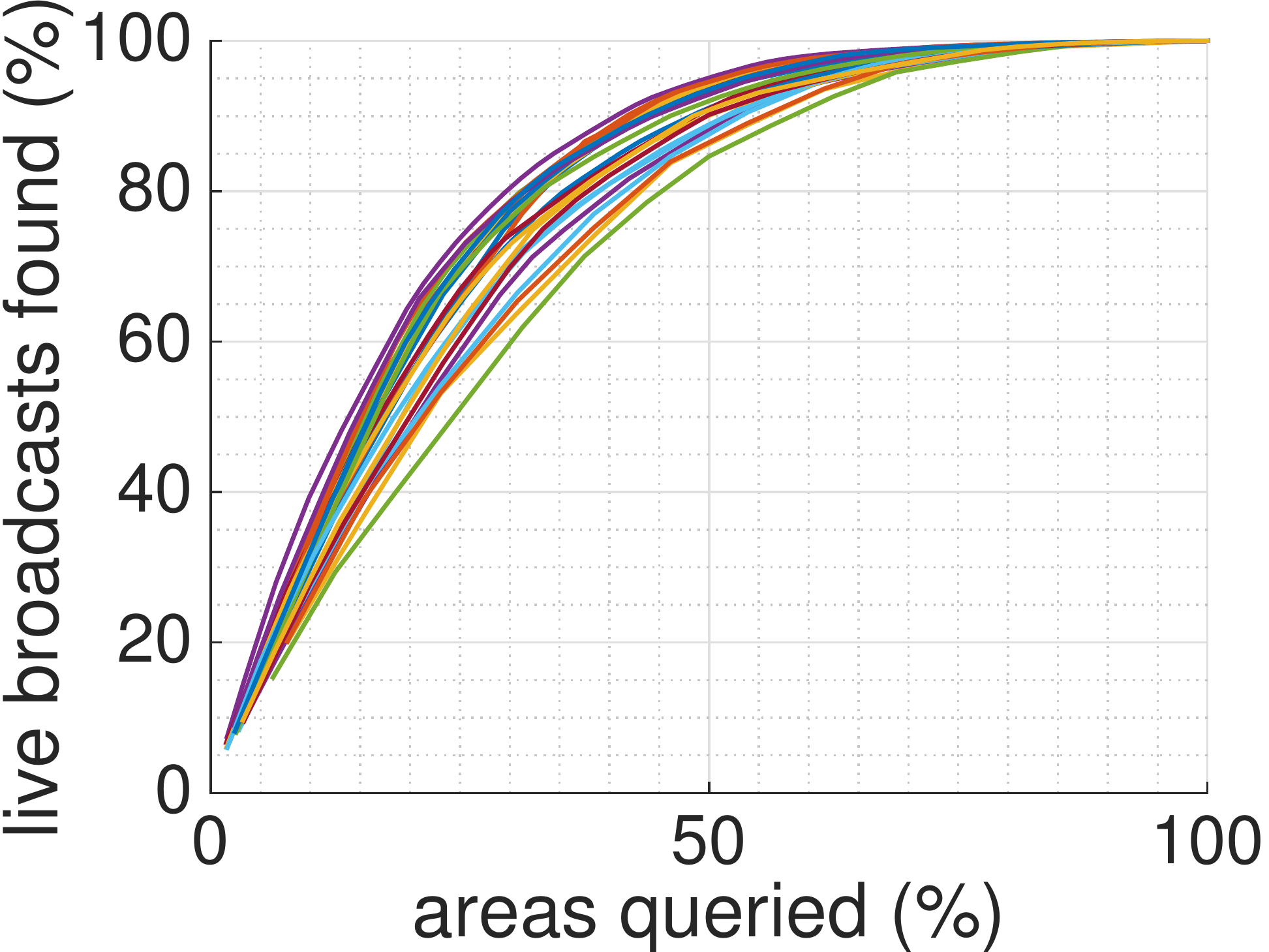}}
\caption{Cumulative number of broadcasts discovered as a function of
  crawled areas (\# of requests). Each curve corresponds to a different deep crawl.}
\label{fig:deep_crawl}
\end{center}
\end{figure}

\begin{figure}[!t]
  \begin{center}
    \subfigure[duration and viewers]{\includegraphics[width=0.49\linewidth]{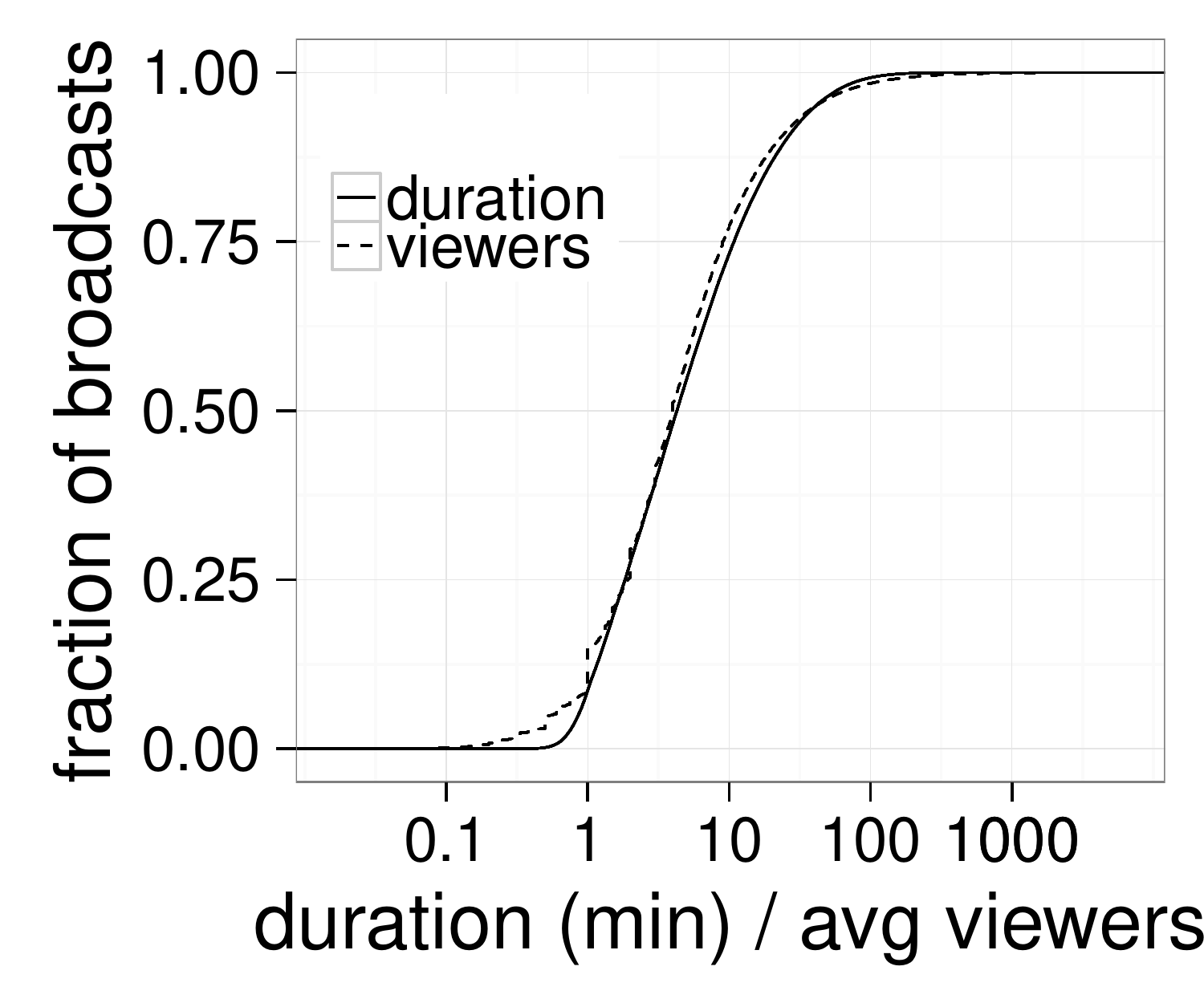}}
    \subfigure[viewers vs. start time]{\label{fig:bc_viewer_hour}\includegraphics[width=0.49\linewidth]{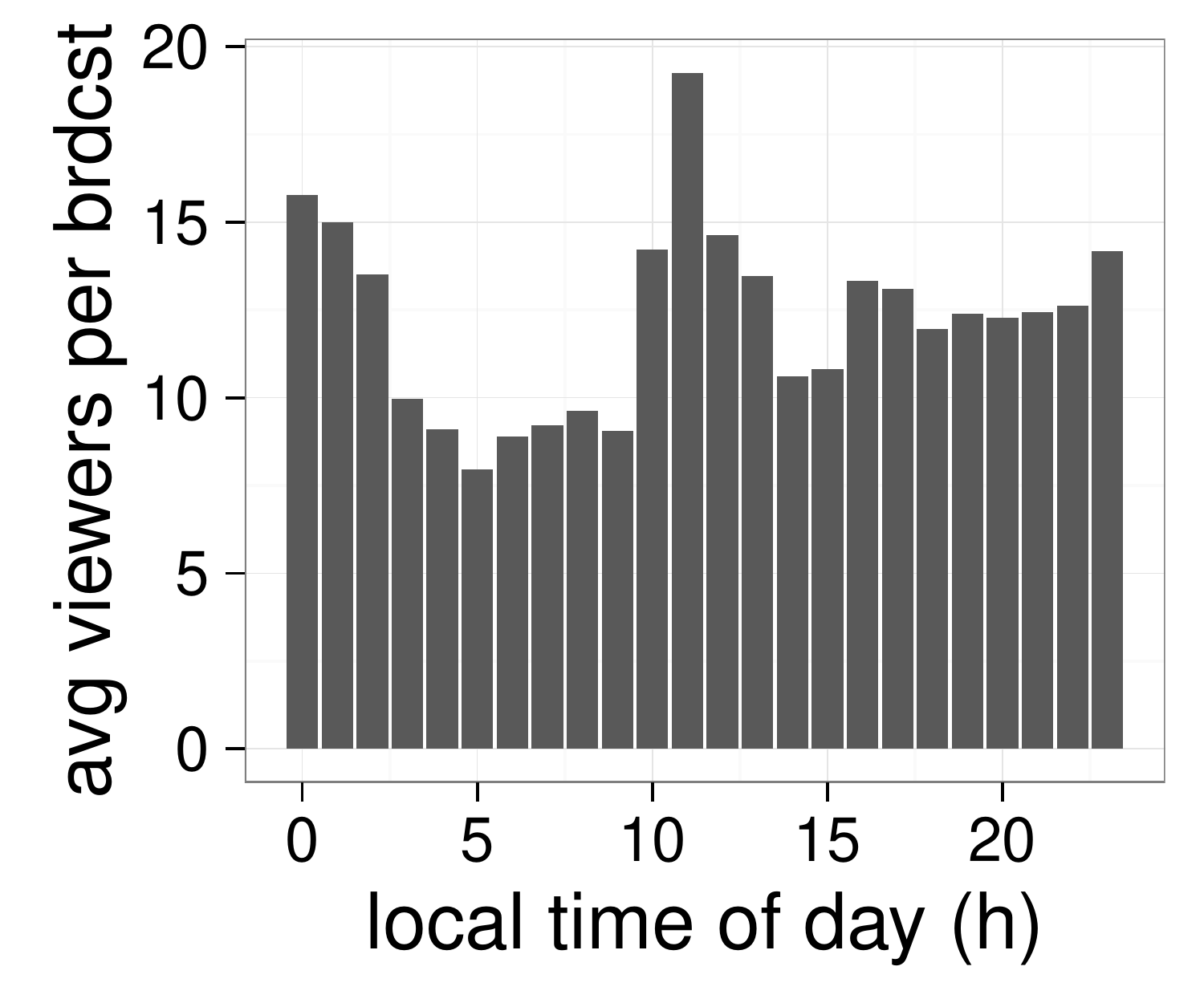}}
    \caption{Broadcasting and viewers.}
    \label{fig:bc_behaviour}
  \end{center}
\end{figure}

Our approach is to first perform a \textit{deep crawl} and then to
select only the most active areas from that crawl and query only them,
i.e., perform a \textit{targeted crawl}. The reason is that deep crawl
alone would produce too coarse grained data about duration and
popularity of broadcasts because it takes over 10 minutes to
finish. In deep crawl, the crawler zooms into each area by dividing it
into four smaller areas and recursively continues doing that until it
no longer discovers substantially more broadcasts.
Such a crawl finds 1K-4K broadcasts\footnote{This number is much smaller
  than the assumed 40K total broadcasts but we miss private broadcasts
  and those with location
  undisclosed.}. 
Figure~\ref{fig:deep_crawl} shows the cumulative number of broadcasts
found as a result of crawls performed at different times of
day. Figure~\ref{fig:deep_crawl_rel} reveals that for all the
different crawls, half of the areas contain at least 80\% of all the
broadcasts discovered in the crawl.
We select those areas from each crawl, 64 areas in total, for a \textit{targeted crawl}.
We divide them into four sets assigned to four different
simultaneously running crawlers, i.e., four emulators running
Periscope with different user logged in (avoids rate limiting) that
repeatedly query the assigned areas. Such targeted crawl completes in
about 50s.

Figure~\ref{fig:bc_behaviour} plots duration and viewing statistics
about four different 4h-10h long targeted crawls started at different
times of the day (note: both variables use the same scale). Broadcast
duration was calculated by subtracting its start time (included in the
description) from the timestamp of the last moment the crawler
discovered the broadcast. Only broadcasts that ended during the crawl
were included (must not have been discovered during the last 60s of a
crawl) totalling to about 220K distinct broadcasts.
Most of the broadcasts last between 1 and 10 minutes and roughly
half are shorter than 4 minutes. The distribution has a long tail with
some broadcasts lasting for over a day.

The crawler gathered viewer information about 134K broadcasts. Over
90\% of broadcasts have less than 20 viewers on average but some
attract thousands of viewers. It would be nice to know the contents of
the most popular broadcasts but the text descriptions are typically not
very informative. Over 10\% of broadcasts have no viewers at all and
over 80\% of them are unavailable for replay afterwards (replay
information is contained in the descriptions we collect about each
broadcast), which means that no one ever saw them. They are typically
much shorter than those that have viewers (avg durations 2min vs. 13
min) although some last for hours. They represent about 2\% of the
total tracked broadcast time. The local time of day shown in
Figure~\ref{fig:bc_viewer_hour} is determined based on the
broadcaster's time zone.
Some viewing patterns are visible, namely a notable slump in the early
hours of the day, a peak in the morning, and an increasing trend
towards midnight, which suggest that broadcasts typically have local
viewers. This makes sense especially from the language preferences
point of view. Besides the difference between broadcasts with and
without any viewers, the popularity is only very weakly correlated
with its duration.

\section{Quality of Experience}
\label{sec:qoe}

In this section, we study the data set generated through automated
viewing with the Android smartphones. It consisted of streaming
sessions with and without bandwidth limit to the Galaxy S3 and S4
devices. We have data of 4615 sessions in total: 1796 RTMP and 1586
HLS sessions without a bandwidth limit and 18-91 sessions for each
specific bandwidth limit. Since the number of recorded sessions is
limited, the results should be taken as indicative. The
fact that our phone had a high-speed non-mobile Internet access means
that typical users may experience worse QoE because of a slower and
more variable Internet access with longer latency.

HLS seems to be used only when a broadcast is very popular.
A comparison of the average number of viewers seen in an RTMP and HLS
session suggests that the boundary number of viewers beyond which HLS
is used is somewhere around 100 viewers. By examining the IP addresses
from which the video was received, we noticed that 87
different Amazon servers were employed to deliver the RTMP streams. We could locate only
nine of them using \texttt{maxmind.com}, but among those nine there
were at least one in each continent, except for Africa, which
indicates that the server is chosen based on the location of the
broadcaster.
All the HLS streams were delivered from only two distinct IP
addresses, which \texttt{maxmind.com} says are located somewhere in
Europe and in San Francisco. We do not currently know how the video
gets embedded into an HLS stream for popular broadcasts but we assume
that the RTMP stream gets possibly transcoded, repackaged, and
delivered to Fastly CDN by Periscope servers. The fact that we used a
single measurement location explains the difference in server
locations observed between the protocols. As confirmed by analysis
in~\cite{wang16imc}, the RTMP server nearest to the broadcasting
device is chosen when the broadcast is initialized, while the Fastly
CDN server is chosen based on the location of the viewing device.

Since we had data from two different devices, we performed a number of
Welch's t-tests in order to understand whether the data sets differ
significantly. 
Only the frame rate differs statistically significantly between the
two datasets. Hence, we combine the data in the following analysis of
video stalling and latency.

\subsection{Playback Smoothness and Latency}
\label{sec:stall_latency}

\begin{figure}[t]
\begin{center}
\subfigure[no bandwidth limiting]{\label{fig:stallratio_rtmp_cdf}\includegraphics[width=0.49\linewidth]{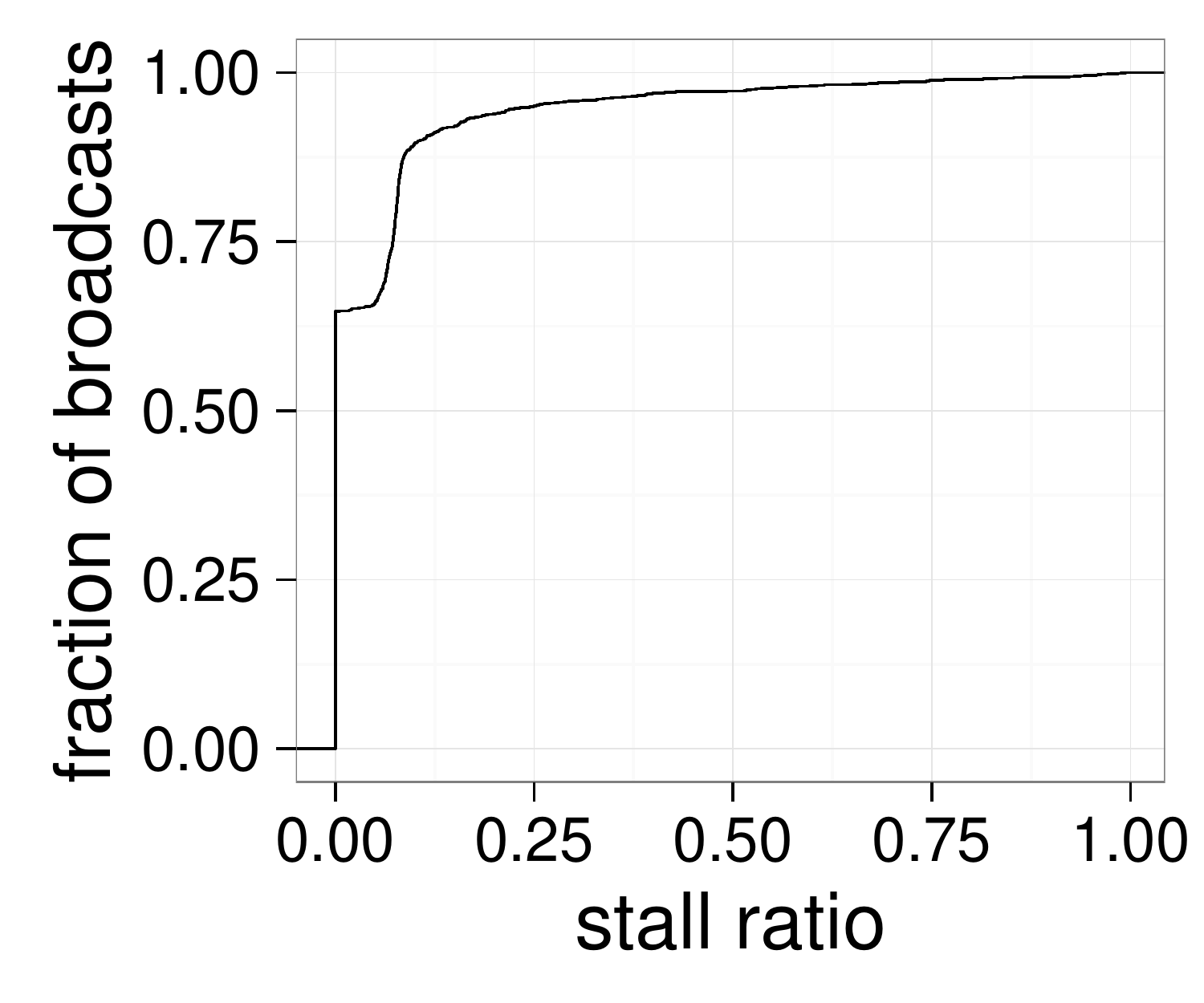}}
\subfigure[bandwidth limiting]{\label{fig:stallratio_rtmp_bw}\includegraphics[width=0.49\linewidth]{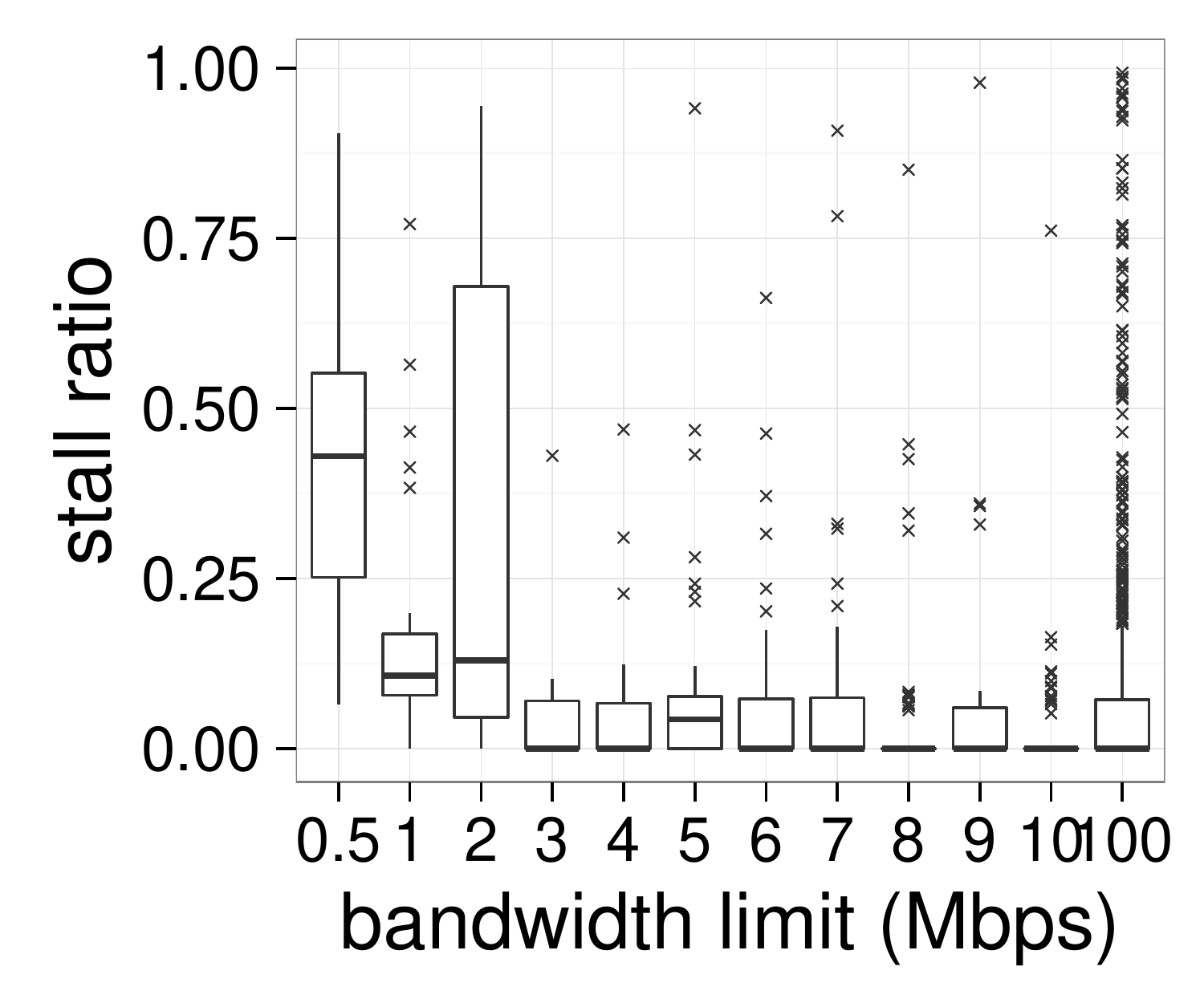}}
\caption{Analysis of the stall ratio for RTMP streams with and without bandwidth limiting.}
\label{fig:stallratio_rtmp}
\end{center}
\end{figure}

We first look at playback stalling. For RTMP streams, the app reports
the number of stall events and the average stall time of an event,
while for HLS it only reports the number of stall events. The
\textit{stall ratio} plotted for the RTMP streams in
Figure~\ref{fig:stallratio_rtmp_cdf} is calculated as summed up stall
time divided by the total stream duration including stall and playback
time. The bandwidth limit 100 in the figure refers to the unlimited
case. Most streams do not stall but there is a notable number of
sessions with stall ratio of 0.05-0.09, which corresponds usually to a
single stall event that lasts roughly 3-5s. The boxplots in
Figure~\ref{fig:stallratio_rtmp_bw} suggest that a vast majority of
the broadcasts are streamed with a bitrate inferior to 2 Mbps because
with access bandwidth greater than that, the broadcasts exhibited very
little stalling. As for the broadcasts streamed using HLS, comparing
their stall count to that of the RTMP streams indicates that stalling
is rarer with HLS than with RTMP, which may be caused by HLS being an
adaptive streaming protocol capable for quality switching on the
fly. 

The average video bitrate is usually between 200 and 400 kbps (see
Section~\ref{sec:audio_video_quality}), which is much less than the 2
Mbps limit we found. The most likely explanation to this discrepancy
is the chat feature. We measured the phone traffic with and without
chat and observed a substantial increase in traffic when the chat was
on. A closer look revealed that the JSON encoded chat messages are
received even when chat is off, but when the chat is on, image
downloads from Amazon S3 servers appear in the traffic. The reason is
that the app downloads profile pictures of chatting users and displays
them next to their messages, which may cause a dramatic increase in
the traffic. For instance, we saw an increase of the aggregate data
rate from roughly 500kbps to 3.5Mbps in one experiment. The precise
effect on traffic depends on the number of chatting users, their
messaging rate, the fraction of them having a profile picture, and the
format and resolution of profile pictures. We also noticed that some
pictures were downloaded multiple times, which indicates that the app
does not cache them.

\begin{figure}[t]
\begin{center}
\subfigure[join time]{\label{fig:jointime_rtmp_bw}\includegraphics[width=0.49\linewidth]{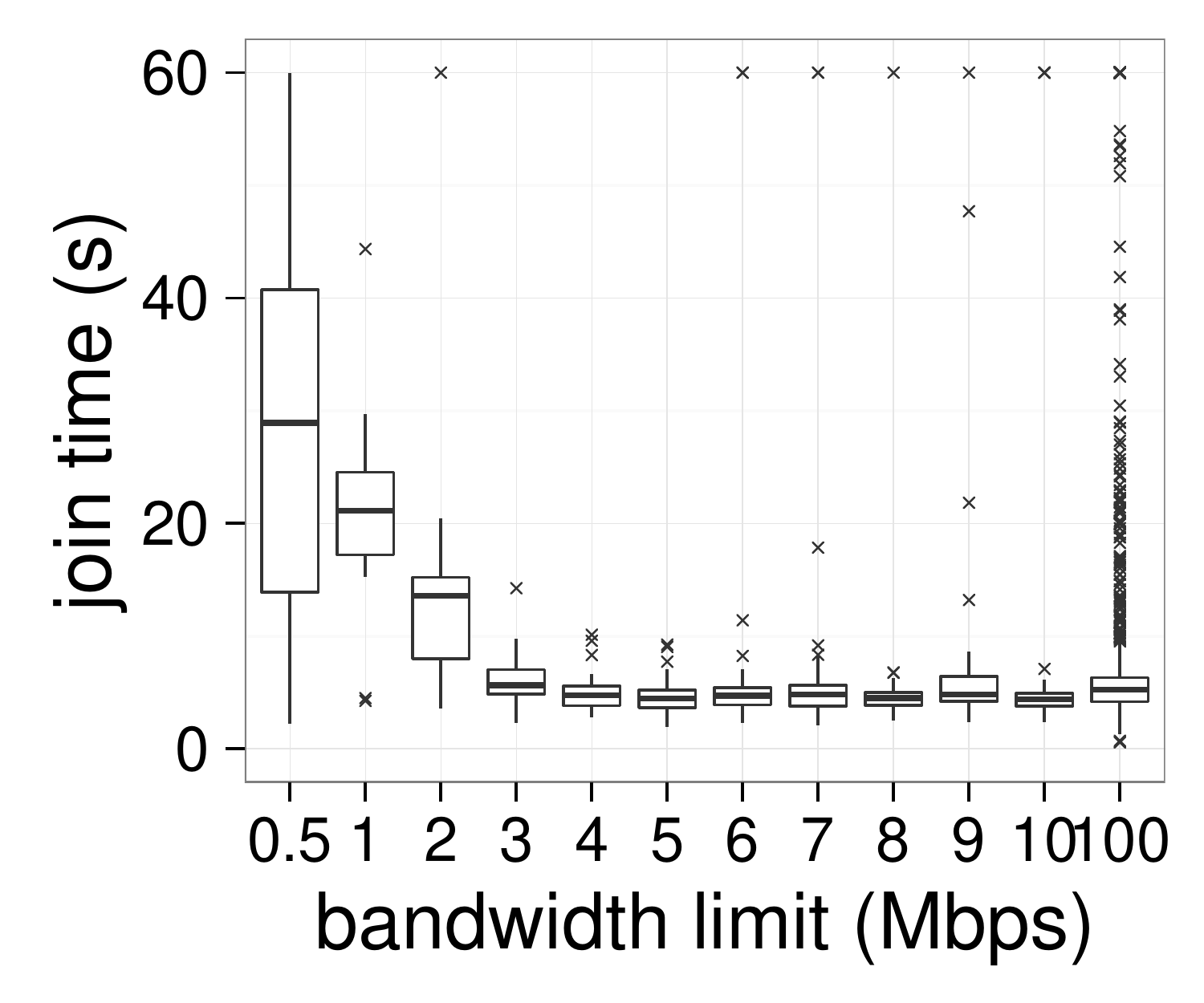}}
\subfigure[playback latency]{\label{fig:latency_rtmp_bw}\includegraphics[width=0.49\linewidth]{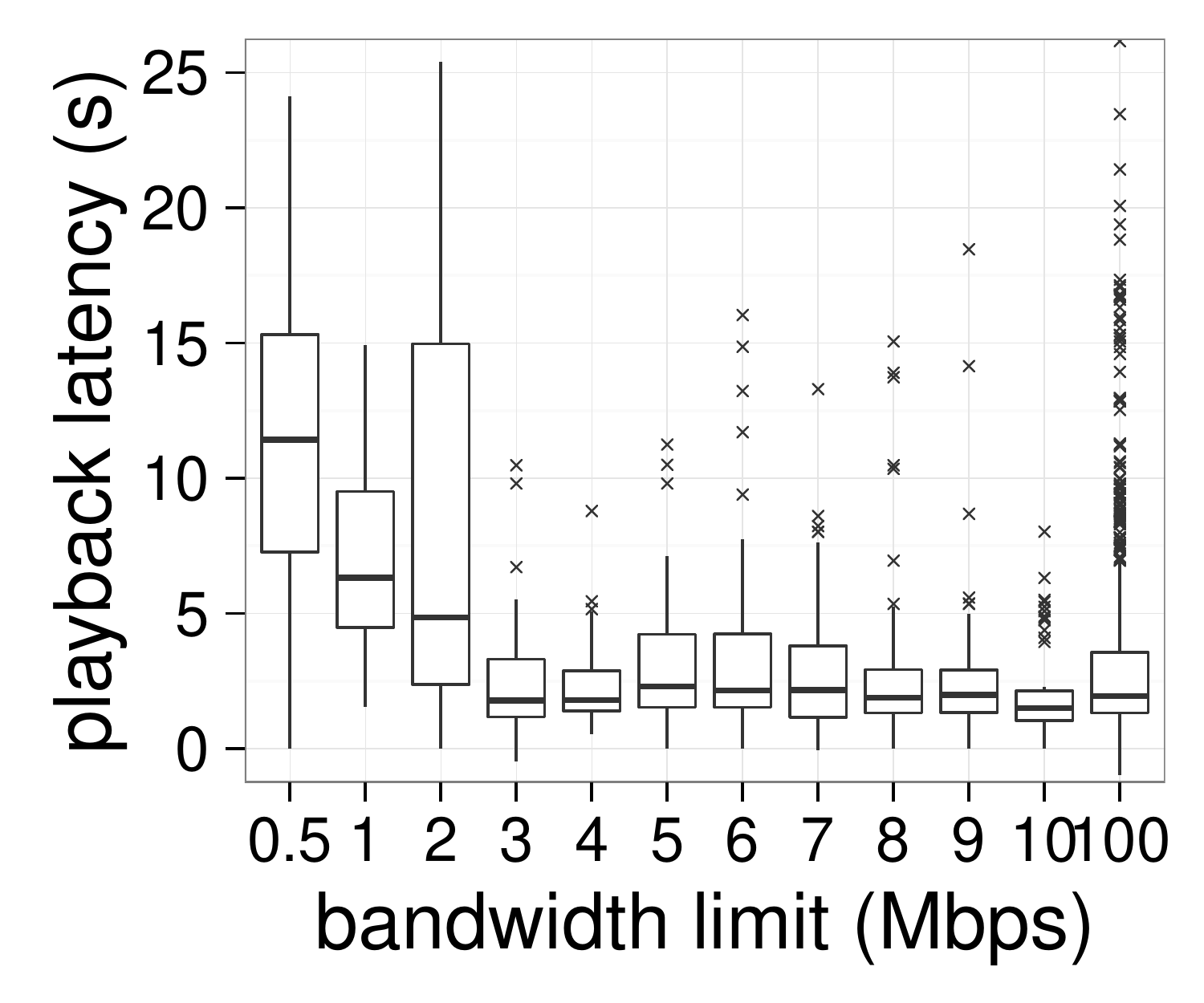}}
\caption{Boxplots showing that playback latency and join time of
  RTMP streams increase when bandwidth is limited. Notice the
  difference in scales.}
\label{fig:jointime_latency}
\end{center}
\end{figure}

Each broadcast was watched for exactly 60s from the moment the
Teleport button was pushed. We calculate the \textit{join time}, often
also called startup latency, by subtracting the summed up playback and
stall time from 60s and plot it in Figure~\ref{fig:jointime_rtmp_bw}
for the RTMP streams. In addition, we plot in
Figure~\ref{fig:latency_rtmp_bw} the playback latency, which is
equivalent to the end-to-end latency. The y-axis scale was cut leaving
out some outliers that ranged up to 4min in the case of playback
latency. Both increase when bandwidth is limited. In particular, join
time grows dramatically when bandwidth drops to 2Mbps and below. The
average playback latency was roughly a few seconds when the bandwidth
was not limited.

\begin{figure}[!h]
  \begin{center}
    \includegraphics[width=0.35\textwidth]{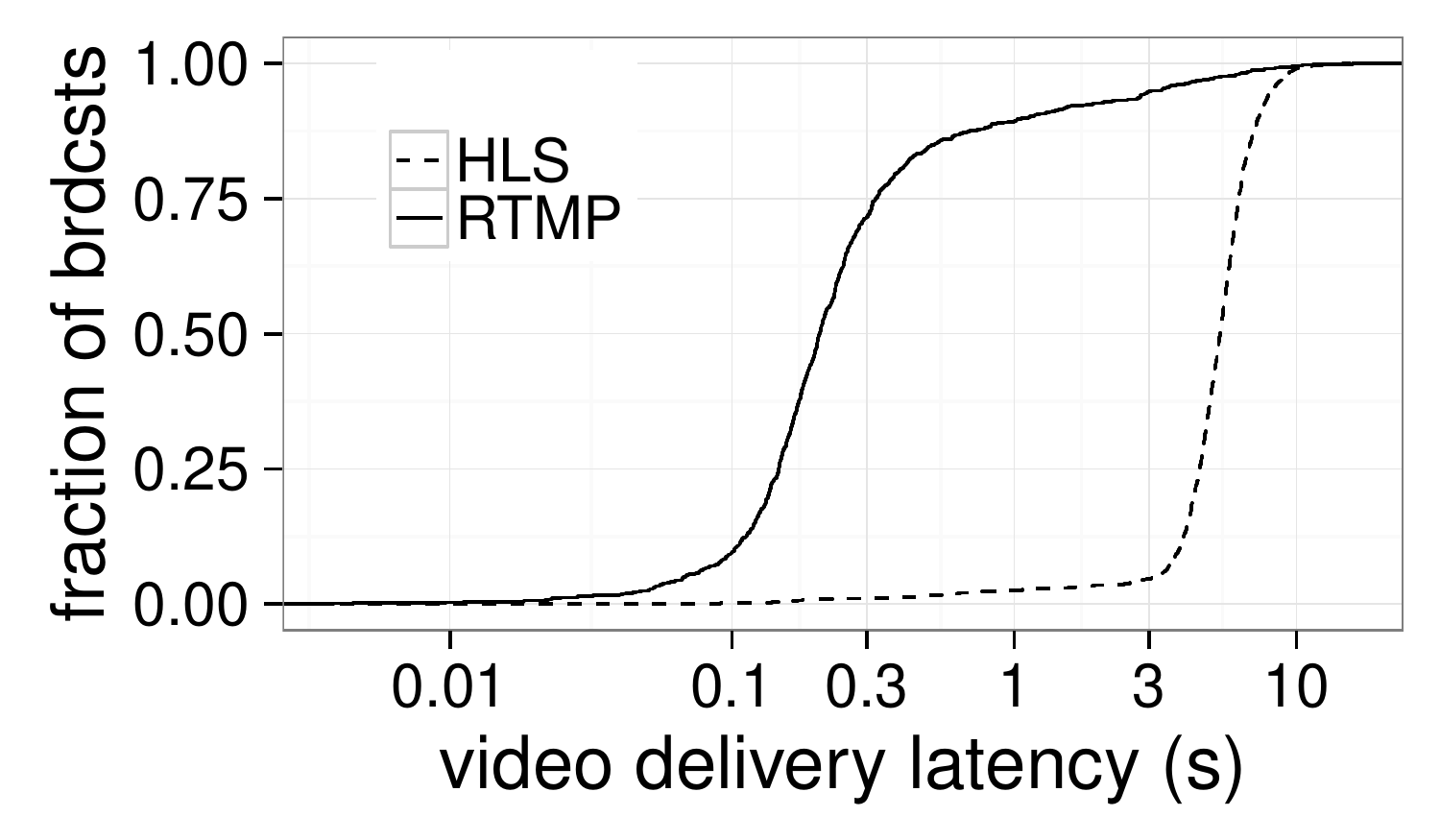}
    \caption{Video delivery latency is much longer with HSL compared
      to RTMP.}
    \label{fig:delivery_latency_ntp}
  \end{center}
\end{figure}

Through experiments where we controlled both the broadcasting and
receiving client and captured both devices traffic, we noticed that
the broadcasting client application regularly embeds an NTP timestamp
into the video data, which is subsequently received by each viewing
client. The experiments indicated that the NTP timestamps transmitted
by the broadcasting device is very close to the tcpdump
timestamps in a trace captured by a tethering machine. Hence, the
timestamps enable calculating the \textit{delivery latency} by
subtracting the NTP timestamp value from the time of receiving the
packet containing it, also for the HLS sessions for which the playback
metadata does not include it. We calculate the average over all the
latency samples for each broadcast. Figure
\ref{fig:delivery_latency_ntp} shows the distribution of the video
delivery latency for the sessions that were not bandwidth limited.
Even if our packet capturing machine was NTP synchronized, we
sometimes observed small negative time differences indicating that the
synchronization was imperfect. Nevertheless, the results clearly
demonstrate the impact of using HLS on the delivery latency. RTMP
stream delivery is very fast happening in less than 300ms for 75\% of
broadcasts on average, which means that the majority of the few
seconds of playback latency with those streams comes from
buffering. In contrast, the delivery latency with HLS streams is over
5s on average. As expected, the delivery latency grows when
bandwidth is limited similarly to the playback latency.
A more detailed analysis of the latency can be found
in~\cite{wang16imc}. The delivery latency we observed matches quite well
with their delay breakdown results (end-to-end delay excluding
buffering).

In summary, HLS appears to be a fallback solution to the RTMP stream. The RTMP
servers can push the video data directly to viewers right after
receiving it from the broadcasting client. HLS delivery requires the
data to be packaged in complete segments, possibly while transcoding
it to multiple qualities, and the client application needs to
separately request for each video segment, which all adds up to the
latency. HLS does produce fewer stall events but we have seen no
evidence of the video bitrate being adapted to the available bandwidth
(Section \ref{sec:audio_video_quality}). It is possible that the
buffer sizing strategy causes the difference in the number of stall
events between the two protocols but we cannot confirm this at the
moment.

\subsection{Audio and Video Quality}
\label{sec:audio_video_quality}

Both RTMP and HLS communications employ standard codecs for audio and video, that is, AAC (Advanced Audio Coding) for audio~\cite{aac_standard} 
and AVC (Advanced Video Coding) for video~\cite{avc_standard}.
In more details, audio is sampled at 44,100 Hz, 16 bit, encoded in Variable Bit Rate (VBR) mode at about either 32 or 64 kbps,
which seems enough to transmit almost any type of audio content (e.g., voice, music, etc.) with the quality expected from capturing through a mobile device.

Video resolution is always 320$\times$568 (or vice versa depending on orientation).
The video frame rate is variable, up to 30 fps. 
Occasionally, some frames are missing hence concealment must be applied to the decoded video.
This is probably due to the fact that the uploading device had some issues, e.g., glitches in the real-time encoding or during upload.

\begin{figure}[t]
\begin{center}
\subfigure[]{\label{fig:bitrate}\includegraphics[width=0.49\linewidth]{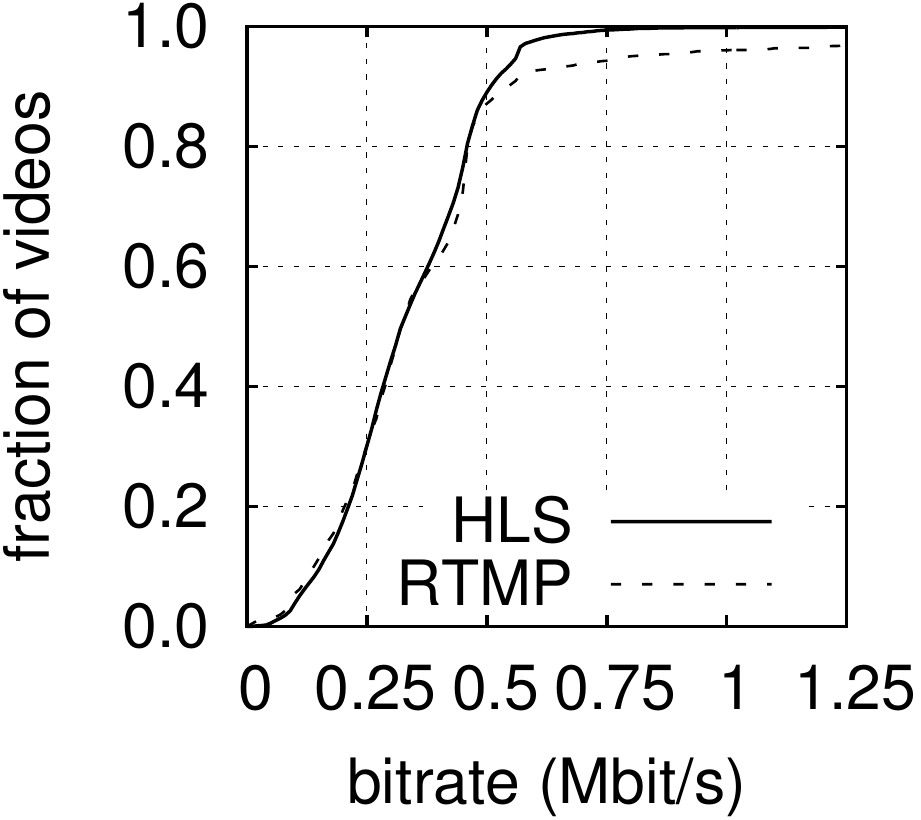}}
\subfigure[]{\label{fig:QP_vs_bitrate}\includegraphics[width=0.49\linewidth]{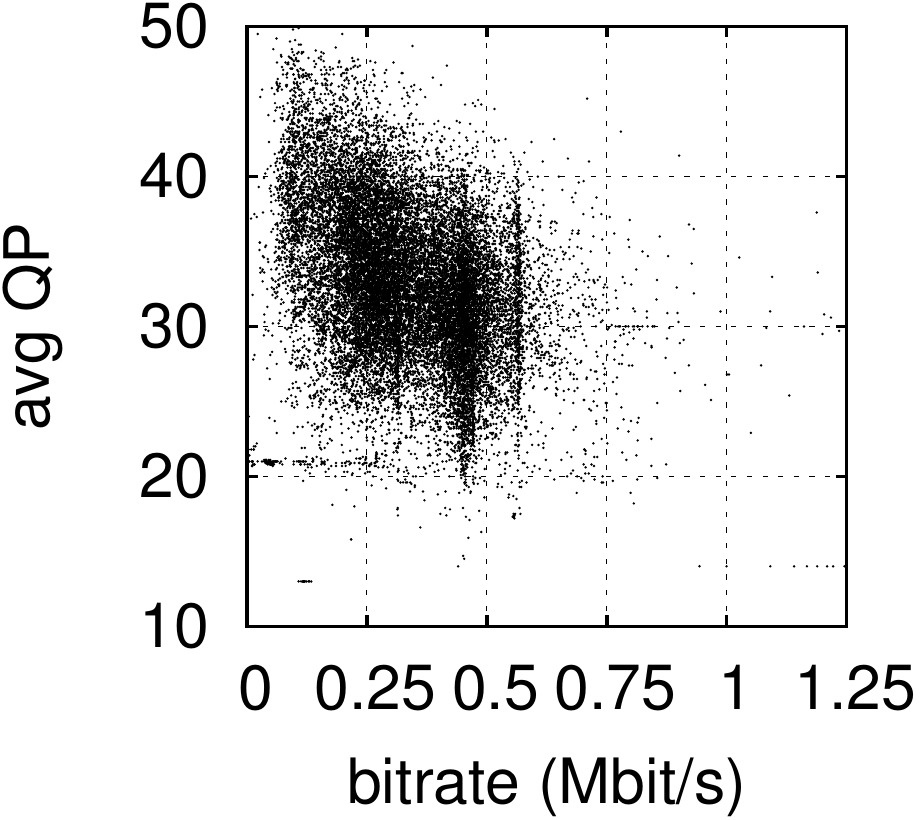}}
\caption{Characteristics of the captured videos.}
\label{fig:segment_duration_bitrate}
\end{center}
\vspace{-7mm}
\end{figure}

Fig.~\ref{fig:bitrate} shows the video bitrate, typically ranging between 200 and 400 kbps. Moreover,
there is almost no difference between HLS and RTMP except for the maximum bitrate which is higher for RTMP.
Analysis of such cases reveals that poor efficiency coding schemes have been used (e.g., I-type frames only).
The most common segment duration with HLS is 3.6 s (60\% of the cases), and it ranges between 3 and 6 s.
However, the corresponding bitrate can vary significantly.
In fact, in real applications rate control algorithms try to keep its average close to a given target,
but this is often challenging as changes in the video content directly influences how difficult is to achieve such bitrate. 
To this aim, the so called quantization parameter (QP) is dynamically adjusted~\cite{chen:ratecontrol_overview07}. 
In short, the QP value determines how many details are discarded during video
compression, hence it can be used as a very rough indication of the quality of
a given video segment.  Note that the higher the QP, the lower the quality and vice versa.

To investigate quality, we extracted the QP and computed its average value for all the videos.
Fig.~\ref{fig:QP_vs_bitrate} shows the QP vs bitrate for each captured video (the whole video for RTMP and each segment for HLS).
When the quality (i.e., QP value) is roughly the same, the bitrate varies in a large range.
On one hand, this is an indication that the type of content strongly differ among the streams. For instance, some of them feature very static content such as one person talking on a static
background while others show, e.g., soccer matches captured from a TV screen.
On the other hand, observing how the bitrate and average QP values vary 
over time may provide interesting indications on
the evolution of the communication, e.g., hints about whether 
representation changes are used.
Unfortunately, we are currently unable to draw definitive conclusions 
since the variation could also be due to significant
changes in the video content which should be analyzed in more depth.

Finally, we investigated the frame type pattern used for
encoding. Most use a repeated IBP scheme. Few encodings (20.0 \% for
RTMP and 18.4\% for HLS) only employ I and P frames only (or just I in
2 cases).  After about 36 frames, a new I frame is inserted. Although
one B frame inserts a delay equal to the duration of the frame itself,
in this case we speculate that the reason they are not present in some
streams could be that some old hardware might not support them for
encoding.

\subsection{Power Consumption}
\label{sec:power}

We connected a Samsung Galaxy S4 4G+ smartphone to a Monsoon Power
Monitor~\cite{msoon} in order to measure its power consumption as
instructed in~\cite{tarkoma14smartphone}.
We used the PowerTool software to record the data measured by the
power monitor and to export it for further analysis. 
 
\begin{figure}[!h]
\begin{center}
\includegraphics[width=\columnwidth]{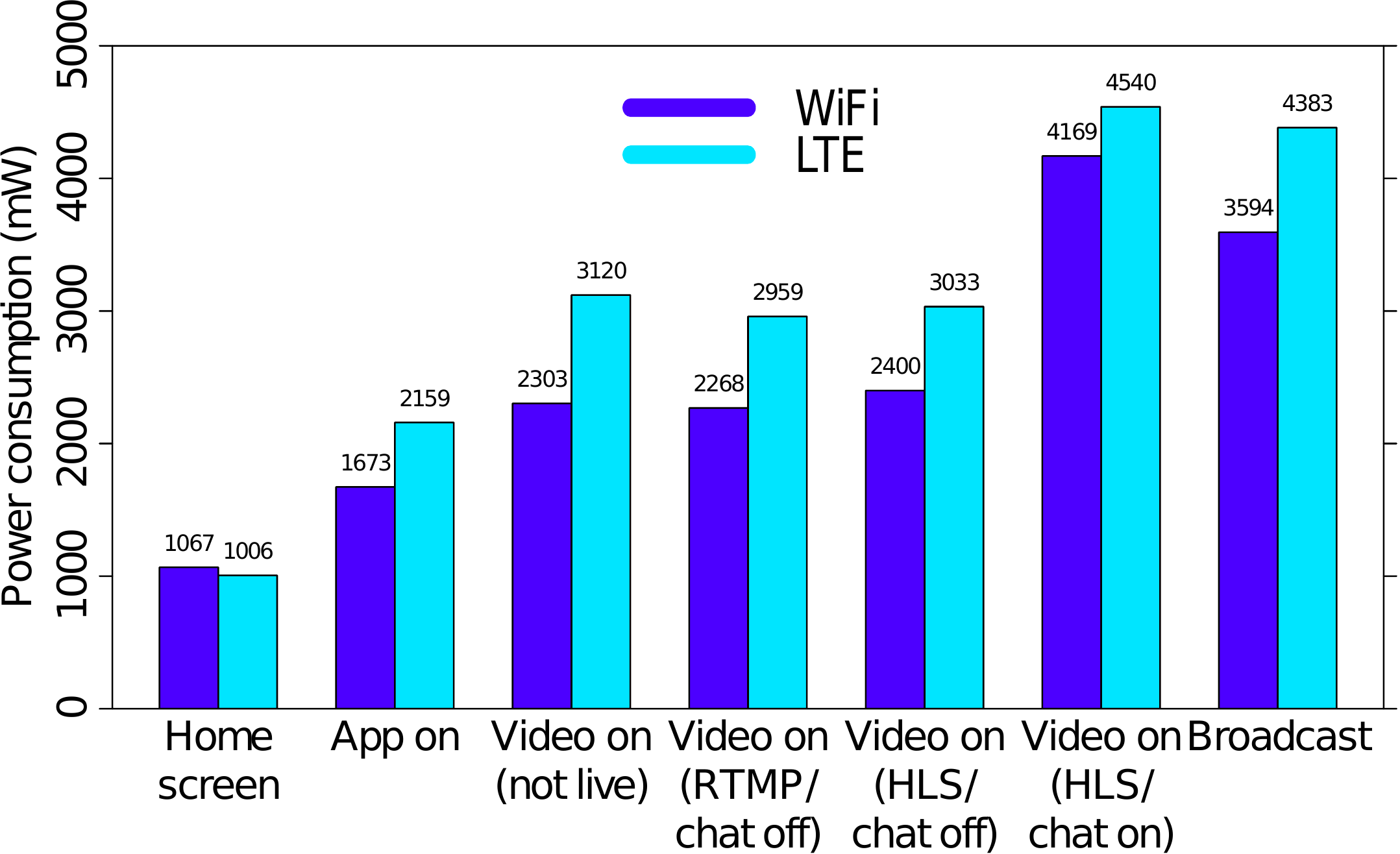}
\caption{Average power consumption with Periscope and idle device.}
\label{fig:power_measurements}
\end{center}
\end{figure}
 
The screen brightness was full in all test cases and the sound was
off. The phone was connected to the Internet through non-commercial
WiFi and LTE networks\footnote{It is a full-fledged LTE network
  operated by Nokia. DRX was enabled with typical timer
  configuration.}. Figure \ref{fig:power_measurements} shows the
results. We measured the idle power draw in the Android application
menu to be around 900 to 1000 mW both with WiFi and LTE
connections. With the Periscope app on without video playback,
the power draw grows already to 1537 mW with WiFi and to
2102 mW with LTE because the application refreshes the available
videos every 5 seconds.
 
Playing back old recorded videos with the application consume an equal
amount of power as playing back live videos. The power consumption
difference of RTMP vs HLS is also very small. Interestingly, enabling
the chat feature of the Periscope videos raises the power consumption
to 2742 mW with WiFi and up to 3599 mW with LTE. This is only slightly
less than when broadcasting from the application. However, the test broadcasts
had no chat displayed on the screen.
 
We further investigated the impact of the chat feature by monitoring
CPU and GPU activity and network traffic. Both processors use DVFS to
scale power draw to dynamic workload~\cite{tarkoma14smartphone}. We
noticed an increase by roughly one third in the average CPU and GPU
clock rates when the chat is enabled, which implies higher power draw
by both processors.
Recall from Section~\ref{sec:stall_latency} that the chat feature may
increase the amount of traffic, especially with streams having an
active chat, which inevitably increases the energy consumed by
wireless communication. The energy overhead of chat could be mitigated
by caching profile pictures and allowing users to disable their
display in the chat.

\section{Related work}
\label{sec:rw}

Live mobile streaming is subject to increasing attention,
including from the sociological point of view~\cite{stewart2016up}.
In the technical domain, research about live streaming focused 
on issues such as
distribution optimization~\cite{lohmar11wowmom}, including scenarios with
direct communication among devices~\cite{zhou16tomm}.
The crowdsourcing of the streaming activity itself also received
particular attention~\cite{he16tmm,zhen16tcsv}.

Stohr et al. have analyzed the YouNow service~\cite{stohr15younow} and
Tang et al. investigated the role of human factors in Meerkat and
Periscope~\cite{tang16chi}.  Little is known, however, about how such
mobile applications perform.  Most of the research has focused on
systems where the mobile device is only the receiver of the live
streaming, like Twitch.Tv~\cite{zhang15nossdav}, or other mobile VoD
systems~\cite{li2012watching_IMC2012}.

We believe that this work together with the work of Wang et
al.~\cite{wang16imc} are the first to provide measurement-based
analyses on the anatomy and performance of a popular mobile live
streaming application. Wang et al. thoroughly study the delay and its
origins but, similar to us, also show results on usage patterns and
video stalling, particularly the impact of buffer size. They also
reveal a particular vulnerability in the service.

\section{Conclusions}
\label{sec:conclusions}

We explored the Periscope service providing insight
on some key performance indicators. Both usage patterns and
technical characteristics of the service (e.g., delay and bandwidth)
were addressed. In addition, the impact of using such
a service on the mobile devices was studied through the
characterization of the energy consumption. We expect that our
findings will contribute to a better understanding of
how the challenges of mobile live streaming
are being tackled in practice.

\section{Acknowledgments}

This work has been financially supported by the Academy of Finland,
grant numbers 278207 and 297892, and the Nokia Center for Advanced
Research.


\end{document}